\newcommand{\fluxhz}{ergs cm$^{-2}$ s$^{-1}$ Hz$^{-1}$}
\newcommand{\ergs}{ergs s$^{-1}$}

\documentclass{aa}   
\usepackage{graphicx}
\usepackage{epstopdf}
\usepackage{epsfig}
\usepackage{soul}
\usepackage[varg]{txfonts}
\usepackage{color}
\usepackage{natbib}
\bibpunct{(}{)}{;}{a}{}{,}
\begin{document}

\title{Environment of $1\le z \le 2$ obscured and unobscured AGNs in the Extended Chandra Deep Field South}
\author{Carlos Guillermo Bornancini\inst{1,2}\fnmsep\thanks{bornancini@oac.unc.edu.ar}  
\and M\'onica Silvia Taormina\inst{2,3} 
\and Diego Garc\'ia Lambas\inst{1,2}}

\authorrunning{Bornancini et al.}
\titlerunning{Environment of obscured and unobscured AGNs} 

\institute{
Instituto de Astronom\'{\i}a Te\'orica y Experimental, (IATE, CONICET-UNC), C\'ordoba, Argentina
\and
Observatorio Astron\'omico, Universidad Nacional de C\'ordoba, Laprida 854, X5000BGR, 
C\'ordoba, Argentina
\and
Nicolaus Copernicus Astronomical Center, Polish Academy of Sciences, ul. Bartycka 18, 00-716 Warszawa, Poland.
}
\date{Received XXX; accepted XXX}
\abstract
{In unified models, different types of active galaxy nuclei (AGN) correspond to a single class of objects, where their observed differences are solely due to the different orientations of the obscuring material around the central inner regions. Recent studies also show that this obscuring material can even extend at galactic scales due to debris from galaxy interactions and/or mergers.
In standard unified models the different AGN types are expected to show similar galaxy environments.} 
{We aim to investigate properties and environment of obscured and unobscured AGNs selected from mid-infrared (MIR) bands from the Multiwavelength Survey by Yale-Chile (MUSYC), in order to test the unified model and evolutionary scenarios.}
{The sample of AGNs was selected from images obtained with the Infrared Array Camera (IRAC) mounted on the Spitzer Space Telescope, based on their MIR colors centered at wavelengths [3.6], [4.5], [5.8] and [8.0] microns. We selected two samples of AGNs with redshifts in the range 1 $\leq$ z $\leq$ 2 and rest-frame absolute magnitudes $M_v\leq -$21: obscured and unobscured AGNs by means of a simple optical-MIR color cut criterion ($R -[4.5] = 3.05$.)} 
{We find that obscured AGNs are intrinsically optically faint in the $R$ band, suggesting that luminous IR-selected AGNs have a significant dust extinction. 
From a cross-correlation with several X-ray surveys, we find that the majority of the AGNs in our sample have X-ray luminosities similar to those found in Seyfert-like galaxies.
We study the properties of galaxies surrounding these two samples. Neighbouring galaxies located close to ($\sim$200 kpc) obscured AGNs tend to have redder colors, compared to the local environment of unobscured AGNs. 
Results obtained from a KS test show that the two color distributions are different at $\sim$95\% confidence level.
We find that obscured AGNs are located in denser local galaxy environments compared to the unobscured AGN sample.}
{Our results suggest that AGN obscuration can occur at galactic scales, possibly due to galaxy interactions or mergers, and that the simple unified model based solely on the local torus orientation may not be sufficient to explain all the observations.}


\maketitle
\section{Introduction}   
The study of high redshift galaxy populations is fundamental in understanding the formation and evolution of galaxies. One of the major difficulties in such studies is the construction of an unbiased and representative sample of galaxies at high-redshift.

Several arguments and pieces of evidence indicate that active galaxy nuclei (AGNs) may play an important role in galaxy formation and evolution. There are two principal arguments: firstly, the relation between star formation history and the AGN activity through cosmic time \citep{madau, bouwens, burgarella, wall}. Secondly, the tight correlation between black hole mass and galaxy bulge mass \citep{ferra} or velocity dispersion \citep{mago}.

There are two popular scenarios for AGNs: unified schemes and evolutionary sequences.
Unified models propose that different AGN types are a single class of physical objects observed under different orientations with respect to the line-of-sight to the observer \citep{urry, antonucci}. In these models it is expected that a large number of objects present a central torus-shaped obscured region (with sizes of a few parsecs) due to large amounts of gas and dust, that can block broad optical spectral lines.

On the other hand, AGN obscuration can also occur at galactic scales. In some models, obscured AGNs represent a phase in the co-evolution of the galaxy an its central black hole. Regarding galaxy mergers, the formation of AGNs exhibit a first phase where the active nuclear region is entirely obscured. After this phase, it follows a short period with infalling material toward the center, with increased star formation activity and galactic winds. Afterwards, the majority of the material is sweeped and the system enters in an unobscured, quasar phase \citep{hop, springel}.

With the advent of highly sensitive infrared detectors such as the Spitzer Space Telescope, it has become clear that optical observations reveal only a small and very biased fraction of the galaxy population \citep{trei04, tre09}.  
Recently, the existence of high redshift objects selected in the mid-infrared, with intrinsically faint optical counterparts has been shown \citep{weed06}. Several studies have related these objects to AGNs obscured by dust that absorbs UV and optical light \citep{houck, hidgon, yan}. 
Color-color diagrams from Spitzer mid-infrared observations using the Infrared Array Camera (IRAC) show a distinct branch in the $[3.6]-[4.5]$ vs. $[5.8]-[8.0]$ bands. \citet{eisen} suggested that point-like sources detected at 3.6 $\mu$m located in this branch were probably AGNs. 
Later studies by \citet{stern05} and \citet{lacy04} have shown that mid-infrared galaxy colors can effectively distinguish AGNs from ``normal'' galaxies. 

\citet{stern05} studied the MIR colors of 800 spectroscopically confirmed AGNs, selected from the AGN and Galaxy Evolution Survey (AGES, \citealt{kocha}) in the Bo\"{o}tes region of the NOAO Deep Wide-Field Survey. From a total of 681 sources residing in this area of the mid-infrared color-color space, Stern et al. found that 522 (77\%) are spectroscopically classified as broad lined AGN, 40 (6\%) are spectroscopically classified as narrow-lined AGN, 113 (17\%) are classified as galaxies, and 6 (1\%) correspond to stars.
\citet{lacy04} using a similar color-color diagram found a wedge-shaped region densely populated by optically selected quasars from the Sloan Digital Sky Survey (SDSS).

In unified models, obscured and unobscured AGNs are expected to reside in similar galaxy environments, and the effect of obscuration is purely an orientation effect.
The properties and evolution of AGNs are also linked to the environment where they reside.
It is well known the association between radio-loud AGNs in massive elliptical galaxies located at the centers of rich clusters of galaxies \citep{hill,best,bornan}.
Several works show that some distant radio sources are embedded in spatially extended ionized gas nebulae of 100$-$200 kpc \citep{vene,miley, villar,roche}, and similar structures are observed in massive ellipticals or cD galaxies at the center of nearby galaxy clusters \citep{west}. 

Recent studies using samples of AGNs at low redshift also suggest that the environment of distinct AGN types are significantly different.
\citet{jiang} analyzed the environmental dependence of a sample of low redshift type 1 and type 2  AGNs taken from the SDSS.
They found that type 2 AGNs and normal galaxies reside in similar environments, but at smaller scales ($\lesssim$ 100 kpc), type 2s have
three times more neighbors than type 1s.
At high redshifts some analyzes suggest that the angular clustering amplitudes for the two types of
AGNs are significantly different, while other studies found no obvious differences. \citet{gilli} studied the spatial clustering of a sample of X-ray selected AGNs at $z\sim1$. They  found no evidence that AGNs with broad optical lines (BLAGN) cluster differently than AGNs without broad optical lines (non-BLAGN). They also found that obscured and unobscured AGNs inhabit similar environments.
\citet{hickox11} presented the first measurement of the spatial clustering of mid-infrared obscured and unobscured quasars in the redshift range $0.7 < z < 1.8$, selected from the 9 deg$^2$ Bo\"{o}tes multiwavelength survey. These authors concluded that obscured AGNs are at least as strongly clustered as the unobscured AGN sample.  
Similar results were obtained by \citet{donoso} who calculated the correlation function of a large sample of AGNs selected from the {\it Wide-Field Infrared Survey Explorer ($WISE$)}. They found that obscured AGNs inhabit denser environments than unobscured ones. In this analysis, obscured AGNs are hosted in dark matter halos with a typical mass of log($M_h$/$M_{\odot}$ $h^{-1}$)$\sim$13.5, while unobscured AGNs inhabit halos of log($M_h$/$M_{\odot}$ $h^{-1}$)$\sim$12.4.   
On the other hand, \citet{dipompeo} calculated the angular clustering of a sample of infrared-selected  obscured and unobscured quasars from the {\it WISE} survey, using the same region as \citet{donoso}. They applied a robust and conservative mask to {\it WISE} quasars and found that obscured quasars reside in halos of mass of log($M_h$/$M_{\odot}$ $h^{-1}$)$=13.3$, while unobscured quasars reside in log($M_h$/$M_{\odot}$ $h^{-1}$)$=$12.8.
These authors found that applying a robust and conservative mask to WISE-selected AGNs yields a weaker but still significant difference in the bias between obscured and unobscured sources.

In this paper we investigate the nature of obscured and unobscured AGNs selected in the mid-infrared, combining optical, MIR and X-ray data, in order to study the environmental dependence of both types of AGNs and their relation to unified models. 

This paper is organized as follows: In Section 2 we present datasets and samples selection, while in Section 3 we investigate about the properties of obscured and unobscured AGN samples. The environment of obscured and unobscured AGNs is analyzed in Section 4. Finally, the conclusions of our work are presented in Section 5. 

Throughout this paper we selected samples of galaxies and AGNs with a limiting magnitude $R_{AB}\leq$ 26.25, as inferred from galaxy number counts.
We use the AB magnitude system\footnote{When necessary the following relations are used:\\IRAC: ([3.6], [4.5], [5.8], [8.0])$_{\rm{AB}}$ = ([3.6], [4.5], [5.8], [8.0])$_{\rm{Vega}}$ + (2.79, 3.26, 3.73, 4.40) \citep{zhu}, $R_{AB}$=$R_{vega}$$+$0.21 \citep{blanton}},  and the same $\Lambda$CDM cosmology adopted by \citet{carda10}, with H$_{0} = 71$ km s$^{-1}$ Mpc$^{-1}$, $\Omega_{M} = 0.3$, $\Omega_{\Lambda} = 0.7$.

\section{Datasets and sample selection}

Observational data were obtained from the MUSYC survey \citep{gawi06, carda10}. MUSYC includes deep images in four areas of 30'x30' and combines an extensive multiwavelength data set, including optical observations in $U$, $B$, $V$, $R$, $I$, $z'$; near IR bands, $J$, $H$ and $K$ and mid-infrared images in [3.6], [4.5], [5.8] y [8.0] microns, obtained from the SIMPLE survey \citep{damen}. It also includes deep optical 18 medium-band photometry from the Subaru telescope \citep{carda10} and rest-frame V-band absolute magnitudes and optical $(U-V)_{rest}$ and $(V-J)_{rest}$ colors. 

In this work we have selected the Extended Chandra Deep Field – South (ECDF-S), which is less affected by dust extinction of the four MUSYC fields. 
The catalog contains 84402 objects with position and flux measurements obtained with the SExtractor package \citep{bertin} and photometric redshifts taken from EazY software using information obtained from 32 bands \citep{brammer}. 
The catalog includes estimates of the quality of each photometric redshift ($Q_z$) which combines the $\chi^2$ of the template fit, the width of the 68\% confidence interval and the Bayesian photometric redshifts (BPZ) odds parameter \citep{benitez}. Values of $Q_z$ increase as any of those parameters deteriorate \citep{carda10, brammer}.
The catalog also provides spectroscopic redshifts which were collected from the literature in order to quantify the accuracy of photometric redshifts. The quality of these redshifts varies widely, from source with multiple spectral lines measurements, to sources that show a simple single spectral line. There are approximately 4000 objects with unique redshifts and $\sim$1000 for which there are multiple measurements. 
Samples 1, 2, and 3 discussed in Subsections 2.1, 2.2, and 2.3 are derived from this parent catalog. 

\subsection{Infrared AGN selection} 

In recent years many authors have presented  several mid-infrared AGN selection criteria. \citet{lacy04} proposes a log(S$_{5.8}$/S$_{3.6}$) vs. log(S$_{8.0}$/S$_{4.5}$), (where $S_{\lambda}$ is the flux density in the corresponding wavelength, in units of microns) color-color diagram and define a region populated by optically selected quasars from the SDSS.    
On the other hand, \citet{stern05} propose a $[3.6]-[4.5]$ versus $[5.8]-[8.0]$ color-color magnitude diagram finding a clear vertical spur populated by bright active galaxies. 
\citet{richards} explored the mid-infrared selection technique of quasars using IRAC observations of SDSS objects. They obtained that the \citet{stern05} selection criteria is significantly less contaminated by inactive galaxies in comparison to that of \citet{lacy04}, although Stern's selection is biased against AGN in the redshift range $3.5 < z < 5$. \citet{richards} also found that at fainter magnitude limits, \citet{lacy04} wedge is contaminated by inactive galaxies.
In order to minimize the contamination of star-forming galaxies observed in the IRAC selection criteria, \citet{donley} proposed a new selection method, which is more restrictive with a smaller selection box than the \citet{lacy04} criterion. 

\citet{lacy13} presented the results of a program of optical and near-infrared spectroscopic followup of AGN candidates selected in the mid-infrared. They found that a significant fraction (22\%) of the objects show no clear AGN signatures in their optical spectra and were classified as ``non-AGN''. Also, some of these galaxies are directly along the AGN locus and satisfy the \citet{donley} selection criteria. These authors also found that the strictness of the \citet{donley} criteria does result in a significant fraction of AGNs being missed.  

\begin{figure}
\includegraphics[width=80mm]{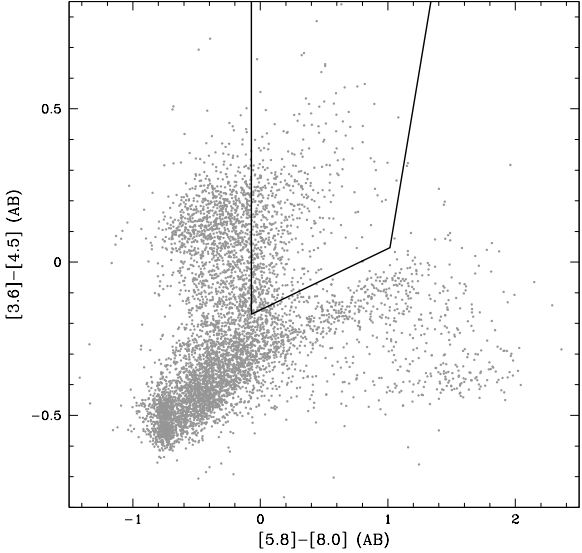}
\caption{Mid-infrared color-color diagram for sources with $\ge$5$\sigma$ detection in all four IRAC bands. The wedge is the AGN selection criterion of \citet{stern05}.}
\label{cometa}
\end{figure}

Following these results, the active galaxy sample selected for this work was identified using the mid-infrared color-color criterion of \citet{stern05}, which is a less strict AGN exclusion cut than that proposed by \citet{donley}.
In Figure \ref{cometa} we plot the $[3.6]-[4.5]$ vs. $[5.8]-[8.0]$ color-color diagram for sources brighter than the 5$\sigma$ detection limits in all IRAC four bands ($[3.6]=23.89$, $[4.5]=23.85$, $[5.8]=22.42$ and $[8.0]=22.50$, \citealt{damen}). Solid lines represent the active galaxy selection criteria of \citet{stern05}.

\begin{figure}
\includegraphics[width=80mm]{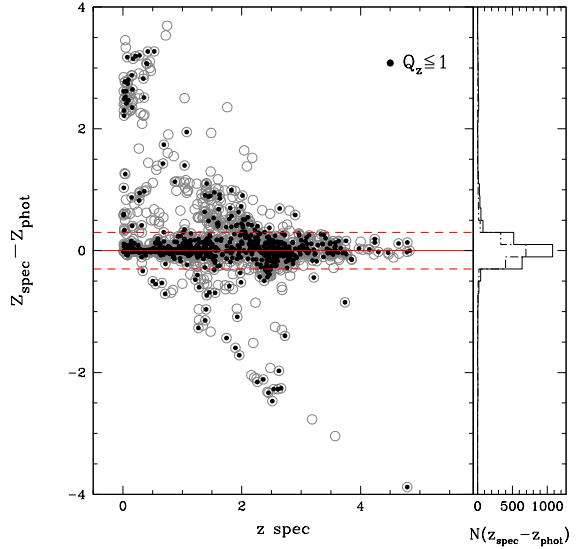}
\caption{Left panel: relation between photometric and spectroscopic redshifts ($z_{spec}-z_{phot}$) as a function of $z_{spec}$. Objects with $Q_z\leq$1 are plotted with black solid dots. Dashed lines represent 1$\sigma$ standard deviation. Right panel: histogram of the difference between spectroscopic and photometric redshift (solid lines) and the corresponding distribution for objects with $Q_z\leq$1 (dashed line histogram).} 
\label{zz_histo}
\end{figure}

In cases where spectroscopic redshifts are not available, we have selected AGNs with high quality ($Q_z\leq1$) photometric redshifts (see \citealt{carda10}).
In Figure \ref{zz_histo} we plot the difference between spectroscopic and photometric redshifts as a function of $z_{spec}$. Filled circles represent objects with high quality photometric redshift, $Q_z\leq$1. As it can be seen, there are objects with bad $z_{spec}$ determinations due to multiple redshift or bad quality measurements. 
When spectroscopic redshifts are available, we have selected those sources with accurate optical spectroscopic redshifts restricted to $|z_{spec}-z_{phot}|\leq0.2$.

\begin{figure*}
\begin{tabular}{cc}
\includegraphics[width=80mm]{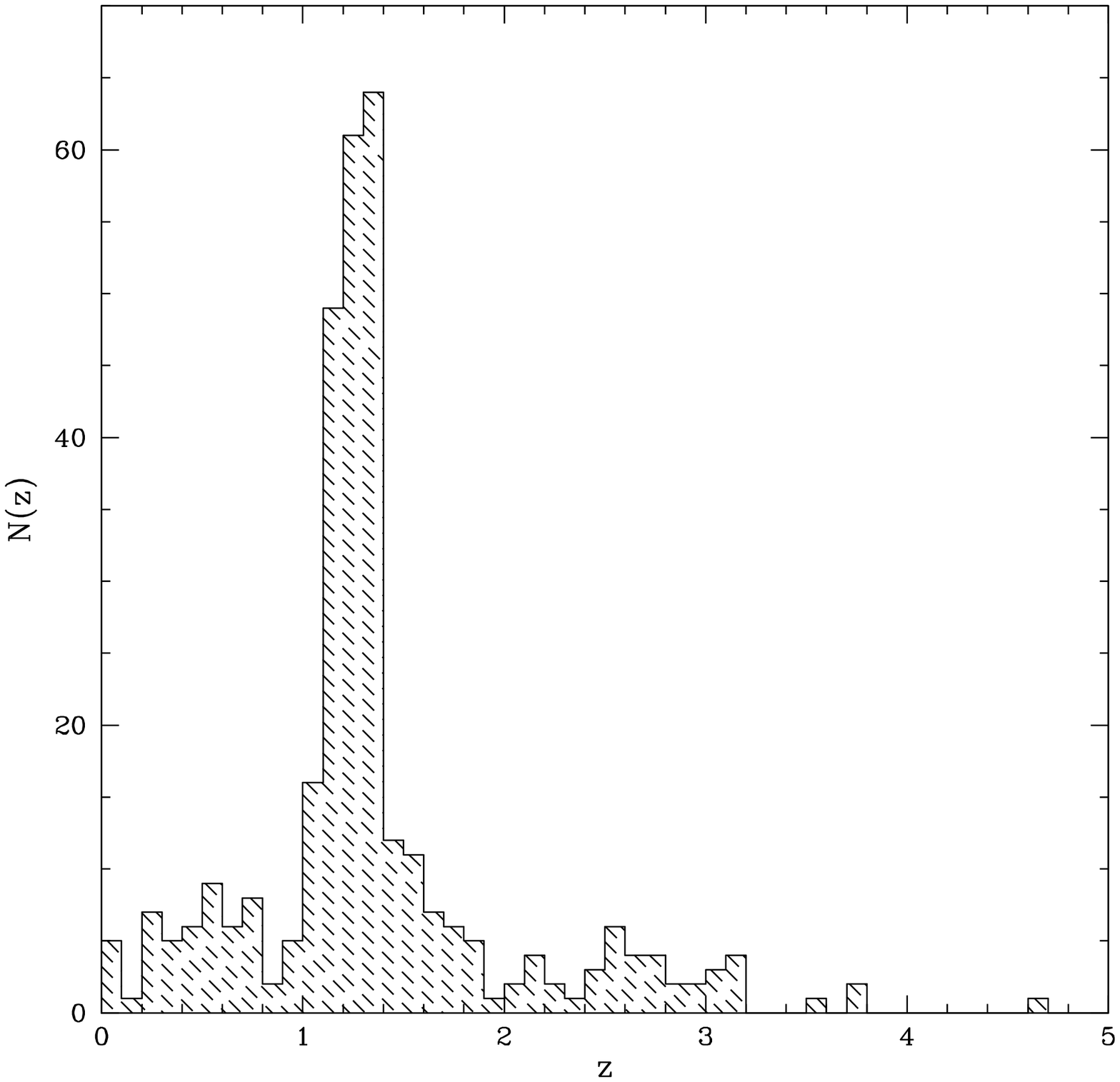}&
\includegraphics[width=80mm]{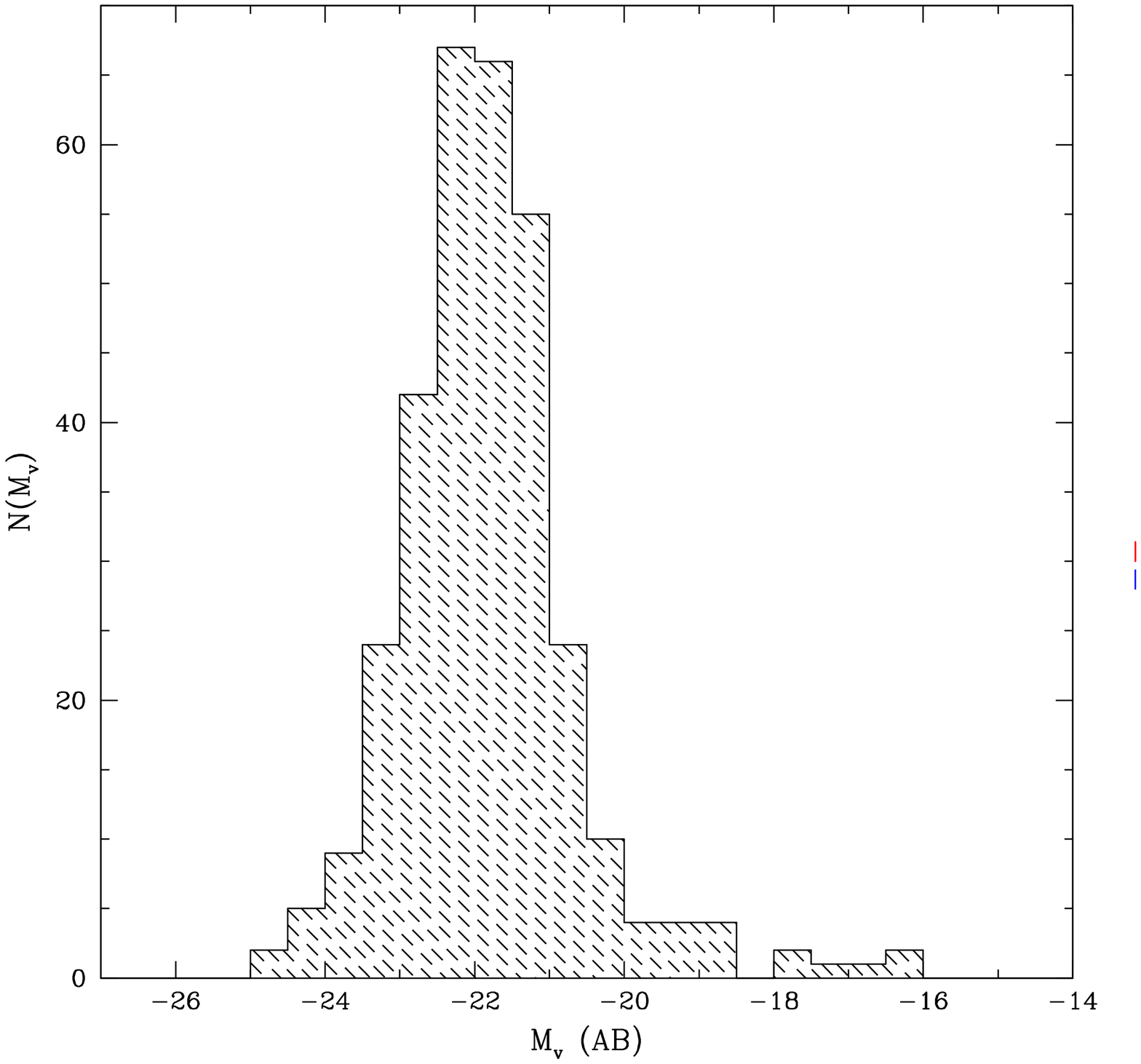}
\end{tabular}
\caption{Left panel: spectroscopic or photometric redshift distribution for AGNs selected using the \citet{stern05} criterion. Right panel: rest-frame absolute magnitudes in V-band distribution of active galaxies selected using the \citet{stern05} criterion}
\label{zzz}
\end{figure*}

In Figure \ref{zzz} (left panel) we show the redshift distribution for objects within the \citet{stern05} wedge, and the distribution of rest-frame absolute magnitudes in the V-band (right panel).
In what follows we will use "Sample 1" to refer to those AGNs that lie in the Stern et al. wedge with $\ge$5$\sigma$ detection in all four IRAC bands and with good spectroscopic or photometric redshift estimates. This AGN sample consists of 327 objects.

\subsection{Obscured and unobscured AGN samples}

\citet{hickox07} noted that the distribution between $R -[4.5]$ color (the optical $R$ band and $[4.5]$ microns in the IR) vs. L$_{4.5 \mu m}$ ($[4.5]$ microns luminosity) is bimodal, and that can distinguish between two different populations, one consists on obscured objects in the optical and the other consists on AGNs with similar properties to type I QSOs (unobscured). They empirically defined the boundary between these two populations at $R -4.5 = 6.04$ (Vega magnitudes). 
Following this work we determined the luminosity in the MIR (4.5 $\mu$m) of each AGN, L$_{4.5 \mu m}$ using,

\begin{equation}
L_\nu(\nu_{\rm rest})=\frac{4\pi d_{\rm L}^2}{1+z} S_\nu(\nu_{\rm obs}),
\end{equation}

where $d_{\rm L}$ is the luminosity distance for a given redshift, $S_\nu$ is the flux density in \fluxhz,
and $\nu_{\rm obs}$ and $\nu_{\rm rest}$ are the observed and rest-frame frequencies respectively, where $\nu_{\rm
rest}=(1+z)\nu_{\rm obs}$. 
We define the luminosity in 4.5 microns in terms of $\nu$$L_\nu$ which unlike the luminosity density $L_\nu$, is not strongly affected by corrections for redshift (see \citealt{hickox07}).

\begin{figure}
\includegraphics[width=80mm]{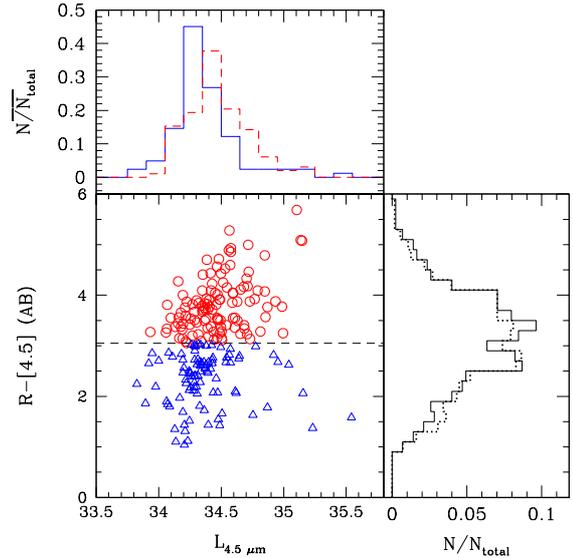}
\caption{Optical-IR color, $R -[4.5]$ vs. L$_{4.5 \mu m}$ ($[4.5]$ microns luminosity for AGN using the \citet{stern05} criterion, with $1\leq z\leq 2$, and $M_v\leq-$21. The horizontal dashed lines show the optical-MIR color-selection criteria for dividing the AGN sample into obscured (open dots) and unobscured (triangles) subsamples. Top panel shows the corresponding distribution of MIR luminosity in 4.5 $\mu$m. Right panel: Solid line histogram shows the color distribution of both AGN samples with $1\leq z\leq 2$, and $M_v\leq-$21. Dotted histogram represents the corresponding distribution for AGNs with $0.7\leq z\leq 3$.}
\label{mari2}
\end{figure}

In Figure \ref{mari2} we plot the $R -[4.5]$ color vs. L$_{4.5 \mu m}$, for the sample of AGNs selected from the \citet{stern05} wedge with $1\le z \le 2$ and $M_v\leq-$21 (hereafter Sample 2).
The criterion for separation of obscured ($R -[4.5] > 3.05$) and unobscured AGNs ($R-[4.5]\leq 3.05$) proposed by \citet{hickox07} using AB magnitudes is shown in the figure.
We identify a sample of 127 obscured and 96 unobscured AGNs. 
In the right panel of this figure we plot the color distribution of AGNs with $1\le z \le 2$, and $M_v\leq-$21. Dotted line histogram represent the corresponding distribution for AGNs with the same redshift cut ($0.7\le z \le 3$) used by \citet{hickox11}. The top panel shows the MIR luminosity distribution in [4.5] microns (L$_{4.5 \mu m}$) for the sample of obscured and unobscured AGNs.  
As we can see, we find similar limits in the optical to MIR color to separate both AGN samples as proposed by these authors. The mean of the distribution is $R-[4.5]=3.07\pm0.07$, which it is indistinguishable from the value proposed by \citet{hickox11}. 
As seen in the left panel of Figure \ref{mari2}, we find no clear evidence of a color bimodality as proposed by \citet{hickox07, hickox11}. These authors restricted the unobscured sample to those spectroscopically identified as broad-line AGNs, to ensure that they unambiguously represent a sample of unobscured quasars.
We find a bimodality in the color distribution when we select objects with $1\leq z\leq 2$, and $M_v\leq-$21 (right panel of Figure \ref{mari2}). 

\subsection{Contamination}

The redshift range of the AGN sample ($1\leq z \leq 2$) was chosen for two reasons. As noted by \citet{hickox07}, AGNs selected in the MIR with $z<0.7$ are possibly heavily obscured normal or star-forming galaxies. For this reason, we exclude objects with $z<1$, in order to minimize contamination by non-AGN objects. In a similar way, we have omitted objects with $M_v>-$21 for the purpose of excluding low luminosity normal galaxies.   
The upper limit in the redshift range was selected in order to exclude high redshift ($z > 2$) star-forming galaxies and also to avoid evolutionary effects between moderately low redshift galaxies ($z\sim0.7$) and those located at higher redshifts ($z>>2$). 

\begin{figure}
\includegraphics[width=80mm]{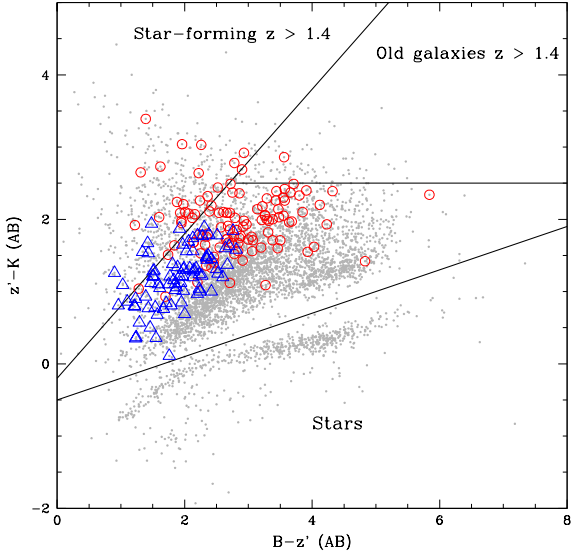}
\caption{Color-magnitude diagram illustrating the selection criteria for the BzK samples, showing the sBzK (star-forming galaxies) and pBzK (passive old galaxies) regions and the location of galactic stars. 
Open circles and triangles represent obscured and unobscured AGN samples, respectively.}
\label{bzk}
\end{figure}

In principle, MIR selected samples do not discriminate between dust heated by an extremely dense hot star-burst and a luminous AGN.
As noted by \citet{donley} the IRAC AGN selection wedges are usually contaminated by star-forming galaxies, especially at high redshift.
In order to decontaminate our AGN samples, we use the BzK diagram \citep{daddi}. 
The BzK selection technique was designed to identify star-forming (sBzK) and passive (pBzK) galaxies at $z >1.4$.
Since the filter set is slightly different than that used by \citet{daddi}, we apply
the offsets determined by \citet{blanc}. The $z-K$ magnitude has been adjusted in $-0.04$ and the $B-z$ magnitude by $+0.56$ to match the filter set used by \citet{daddi}. We plot in Figure \ref{bzk} the color-color $z'-K$ vs. $B-z'$ diagram for all sources with $K_{AB}\le 21.84$ (i.e $K_{vega}\leq 20$, which is the limiting magnitude used by \citealt{daddi}).

We also plot the corresponding colors ($z'-K$ vs. $B-z'$) of obscured and unobscured AGN samples. We find that there are some obscured and unobscured objects in a region populated by star-forming galaxies. In order to selected a sample of pure AGNs, we have removed these objects from our final sample. 

In Figure 6 we show the final color-luminosity distribution of pure AGNs (same as Sample 2 but without the contamination of star-forming galaxies detected from the BzK diagram). This final sample consists of 105 obscured and 86 unobscured AGNs (Sample 3). 

In order to quantify the presence of a color bimodality, we perform a Gaussian mixture modeling (GMM) test.
We have used the GMM code of \citet{muratov} to quantify the probability that the color distributions are better described by a bimodal rather than a unimodal distribution.
This code uses information from three different statistic tools: the kurtosis, the distance from the mean peaks ($D$), and the likelihood ratio test (LRT).
The kurtosis test measures the degree of peakedness of a distribution: a positive value corresponds to a sharply
peaked distribution whereas a negative kurtosis corresponds to a flattened distribution.
The distance from the mean peaks or ``Bandwidth test'', is the separation between the means of the Gaussian components relative to their widths, calculated as
\begin{equation}
D=\frac{\big|\mu_1-\mu_2 \big|}{\sqrt{\frac{\sigma^2_1+\sigma^2_2}{2}}}
,\end{equation}

where $\mu_1$ and $\mu_2$ are the mean values of the two peaks of the proposed bimodal distribution, and $\sigma_1$, $\sigma_2$ are the corresponding standard deviations. For a clear separation between the two peaks it is required that $D > 2$.

LRT is defined as

\begin{equation}
LRT=2\times\left[ln(L_{bimodal})-ln(L_{unimodal})\right]
,\end{equation}

\begin{figure}
\includegraphics[width=80mm]{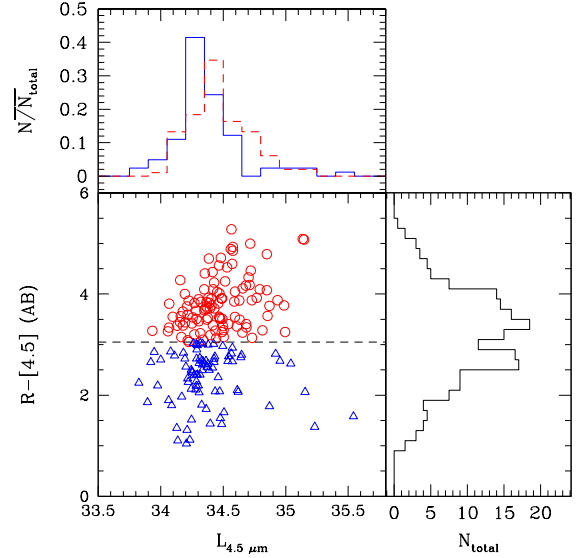}
\caption{Same as Figure \ref{mari2}, but excluding star-forming galaxies selected from the BzK diagram.}
\label{marisin}
\end{figure}

where $L_{bimodal}$ and $L_{unimodal}$ are the likelihood for a bimodal and unimodal distributions, respectively.

The GMM code provides the probability $p(\chi^2)$ of rejection of a unimodal distribution in favour of a bimodal fit. 

This program requires approximate initial values for the two estimated peaks in the observed distribution.
In order to reach a maximun value of $p(\chi^2)$ we have performed tests with different values of $\mu_1$ and $\mu_2$.
For the sample of AGNs selected from \citet{stern05} wedge with the same redshift range ($0.7\le z \le 3$) used by \citet{hickox11}, we find $\mu_1=1.8\pm0.4$ and $\mu_2=3.3\pm0.3$, with $p(\chi^2)=0.05$, $D=2.5\pm0.5$, kurtosis$=-0.45$ (see the right panel of Figure \ref{mari2}).
For AGNs with $1\leq z\leq 2$, and $M_v\leq-$21 (Sample 2), we obtained $\mu_1=2.2\pm0.6$ and $\mu_2=3.5\pm0.6$, with $p(\chi^2)=0.44$, $D=2.6\pm0.7$, kurtosis$=-0.30$ (see the right panel of Figure \ref{mari2}).
For pure AGNs (see Sample 3), we find $\mu_1=2.1\pm0.6$ and $\mu_2=3.5\pm0.3$, with $p(\chi^2)=0.60$, $D=2.2\pm0.7$, kurtosis$=-0.40$, (see the right panel of Figure \ref{marisin}). 
In order to estimate the reliability of the results, we have performed tests using randomly selected subsamples with half the total number of AGNs. For a set of these subsamples  we have calculated the corresponding parameters using the GMM code in a similar fashion as in the total sample.
We obtained similar parameters to those of the full sample: negative kurtosis,  $D>2$, and $p(\chi^2)\sim$60 to 90\%, showing the stability of the results in spite of the low number statistics.
This analysis shows that the color distribution of pure AGNs with $1\leq z\leq 2$, and $M_v\leq-$21 (Sample 3) is bimodal, as in previous works by \citet{hickox07, hickox11}. 

\section{Properties of obscured and unobscured galaxy AGN samples} 

\begin{figure}
\includegraphics[width=80mm]{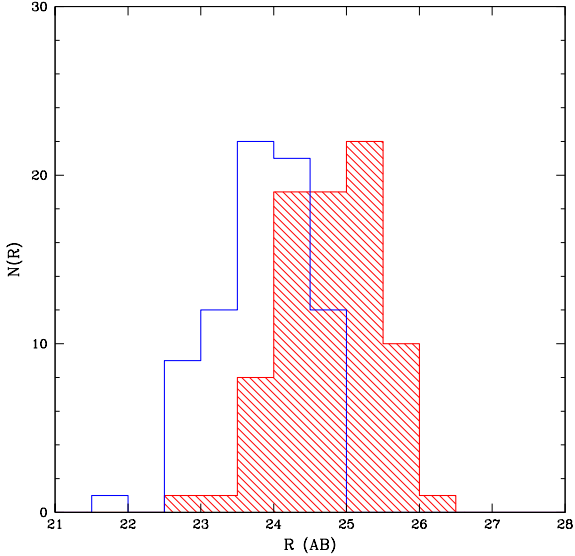}
\caption{R-band distribution for obscured (shaded histogram) and unobscured (solid line histogram) AGN samples.}
\label{R}
\end{figure}

In Figure \ref{R} we plot the distribution of R-band apparent magnitudes for both AGN samples. 
As in previous results \citep{hickox07}, we find that obscured objects are inherently faint, possibly due to the presence of large amounts of dust that absorb UV radiation from both optical and nuclear region. We analyze the SExtractor stellaricity parameter {\tt CLASS\_STAR}, which indicates the likelihood of an object to be a galaxy or a star based on a neuronal network technique \citep{bertin}. In the ideal case, a galaxy and a star have a stellaricity index {\tt CLASS\_STAR} $= 0.0$ and $1.0$, respectively.
We find that most of obscured galaxies have an extended emission, while unobscured AGNs have a mixture of extended and point-like emission. One hundred percent (100\%) of obscured galaxies have {\tt CLASS\_STAR}$<0.5$, suggesting that their optical emission is dominated by the host galaxy, while unobscured sample have 13\% of galaxies with {\tt CLASS\_STAR}$>0.7$ and 87\% have {\tt CLASS\_STAR}$<0.7$.


 
\subsection{X-Ray properties} 

In order to explore the X-ray properties of obscured and unobscured galaxy samples, we select X-ray AGN candidates. Several X-ray surveys have overlapping coverage of the same region of study: the 1, 2 and 4 Ms Chandra Deep Field-South (CDF-S \citealt{gia,A03,luo08,xue}), and the deep ECDF-S \citep{V06}. 
We cross-correlated the spatial positions of objects detected in all four IRAC bands with their X-ray source positions from a combined catalog from \citet{gia,A03,V06,luo08} and \citet{xue}, using a matching radius of 2 arcsec. From a total of 1396 X-ray sources, we find 1173 (84\%) mid-infrared matches.
We determined the hardness ratio (which is an indication of the X-ray spectral shape) of each X-ray source using the following relation,

\begin{figure}
\includegraphics[width=80mm,angle=0]{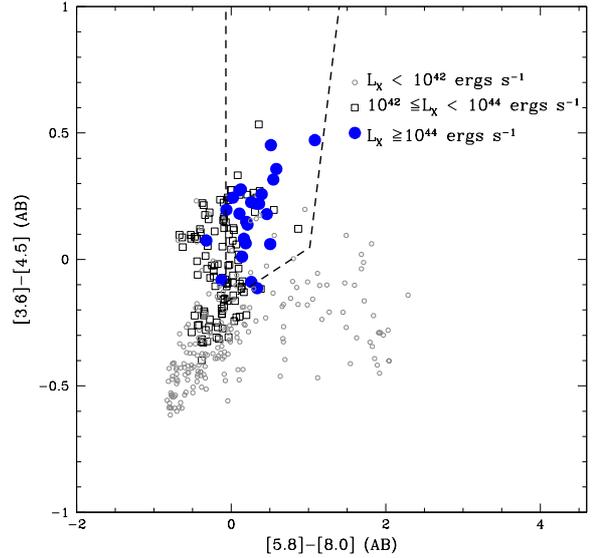}
\caption{Mid-infrared color-color diagram for X-ray emitting galaxies. Open circles represent objects with $L_X<10^{42}$\ergs (normal galaxies), open squares are the values obtained for Seyfert-like AGNs with $10^{42}\leq L_X<10^{44}$ \ergs and filled dots are the corresponding values for QSOs with $L_X\geq10^{44}$ \ergs. The wedge is the AGN selection criterion of \citet{stern05}.}
\label{cometax}
\end{figure}

\begin{figure}
\includegraphics[width=80mm,angle=0]{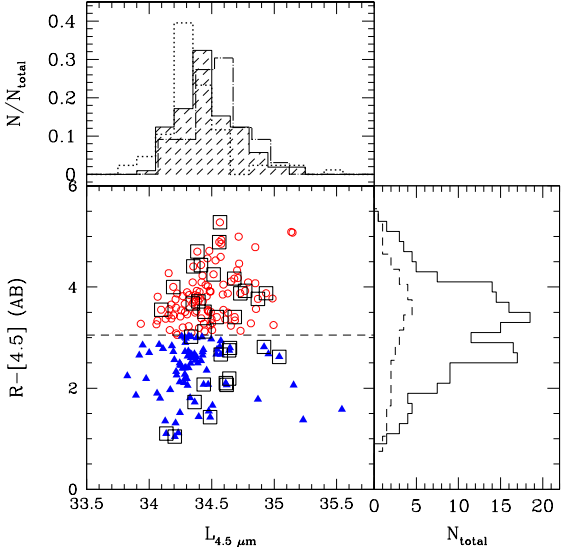}
\caption{Optical-IR color, $R-[4.5]$ vs. L$_{4.5 \mu m}$. X-ray emitting AGNs are plotted with open squares. 
Top panel shows the corresponding distribution of MIR luminosity in the 4.5 $\mu$m for obscured (shade histogram), unobscured (dotted line) AGNs, and X-ray emitting AGNs (dot-long dash histogram). We introduce an artificial shift of $0.02$ in the dot-long dash histogram for clarity. Right panel shows the color distribution of both AGN samples (solid line histogram) and the corresponding distribution considering AGN with X-ray emission (dash lines).}
\label{marix}
\end{figure}

\begin{equation}
{\rm HR}=\frac{H-S}{H+S},
\end{equation}
where $H$ and $S$ are the count rates in the hard ($2-8$ keV) and soft ($0.5-2$ keV) bands, respectively. 
For all the X-ray sources in our matched catalogs, we also computed the rest-frame X-ray luminosity calculated as, 

\begin{equation} 
L_{X} = 4 \pi\,d^{2}_{L}\,f_{X}\,(1+z)^{\Gamma-2}\,\rm{erg}\,\rm{s}^{-1}, 
\end{equation} 	
	
where $d_{L}$ is the luminosity distance (cm), \emph{f}$_{X}$ is the X-ray flux (erg s$^{-1}$ cm$^{-2}$) in the full-band ($0.5-8$ keV) and the photon index was assumed to be $\Gamma=1.8$ \citep{Tozzi06}. The luminosity distance was calculated using either the spectroscopic redshift or, if not available, the photometric redshift estimate.

In Figure \ref{cometax} we plot MIR colors for X-ray emitting normal galaxies with $L_X<10^{42}$ \ergs, Seyfert-like galaxies; $10^{42}\leq L_X<10^{44}$ \ergs and for QSOs with $L_X\geq10^{44}$ \ergs. As it can be seen, the majority of the X-ray emitting QSOs fall into the region proposed by \citet{stern05}, Seyfert-like galaxies display a wide range of MIR color distribution and X-ray selected normal galaxies are mainly located in the principal sequence (see Figure \ref{cometa}), and only a few of these galaxies fall inside the AGN wedge of \citet{stern05}.

In Figure \ref{marix} we plot optical-MIR color ($R-[4.5]$) vs. L$_{4.5 \mu m}$ as in Figure \ref{marisin}, but including AGNs taken from Sample 3 with detected X-ray emission (Open squares). We find that most of X-ray emitting active objects selected in the MIR have red $R-[4.5]$ colors.
We find 19 obscured and 14 unobscured AGNs with X-ray emission.  

\begin{figure}
\includegraphics[width=80mm]{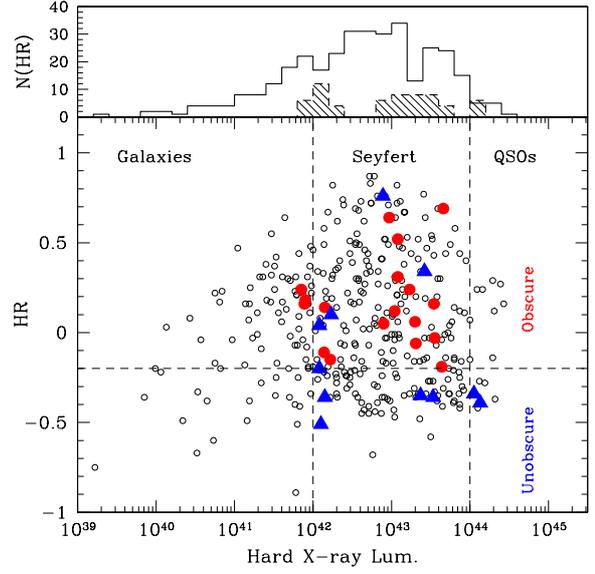}
\caption{Hardness ratio as a function of hard X-ray luminosity for all MIR matches (Open circles). Sources classified as obscured and unobscured sources are plotted with filled circles and triangles, respectively. Vertical dashed lines show the typical separation for normal galaxies, Seyfert-like galaxies and quasars used in the X-rays. The dashed horizontal line shows the HR value for a source with $N_H>10^{21.6}$ cm$^2$ at $z>1$. In the upper panel we plot the hard X-ray luminosity distribution for all the sources (solid line histogram) and dashed histogram represent the corresponding distribution for MIR selected AGNs (obscured and unobscured samples). 
For a better comparison, the AGN hard X-ray luminosity (shade histogram) have been scaled up by a factor of two to show their relative values compared to the whole sample of X-ray source detections.}
\label{hrc}
\end{figure}

Following \citet{treis09} we plot the hardness ratio as a function of hard X-ray luminosity (2$-$8 KeV) (see Figure \ref{hrc}). Open circles represent matches in all four IRAC band. We also plot the corresponding values for the obscured and unobscured X-ray emitting AGNs. The dashed horizontal line shows the HR value (HR$=-0.2$) for a source with a neutral hydrogen column density, $N_H>10^{21.6}$ cm$^2$ at $z>1$ \citep{gilli}. This limit was used in the X-rays as a separation between obscured and unobscured sources.

As it can be seen, sources with hard X-ray spectra (positive hardness ratio) tend to be also classified as obscured sources in the optical, while unobscured AGNs have in general a soft X-ray spectrum (negative hardness ratio). The soft X-ray emission of obscured AGNs tend to be absorbed, while hard X-ray are able to escape.
It is interesting that some obscured and unobscured AGN are located near the boundary between normal and Seyfert-like galaxies, while the majority have X-ray luminosities in the range $10^{43} < L_X < 10^{44}$ \ergs.

\subsection{Galaxy color distribution}

\begin{figure}
\includegraphics[width=80mm]{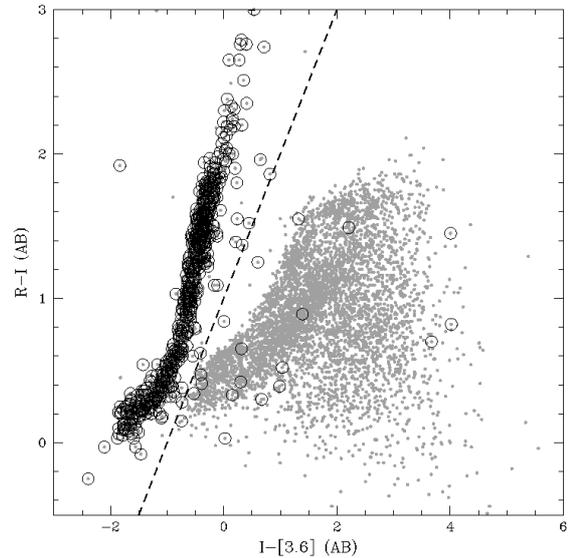}
\caption{Color-color diagram used to remove stars from our final galaxy catalog. Stars have colors that are clearly separated from the
regions occupied by galaxies and can be efficiently isolated with the criterion  $R-I>(I-[3.6])+1$ (red line). Open blue circles represent stellar objects identified in Figure \ref{bzk}.}
\label{remo}
\end{figure}

In this Section, we study the color of neighbor tracer galaxies located in the field of obscured and unobscured AGN samples.
We have applied an apparent color cut method implemented by \citet{brown} to exclude foreground stars from our final catalog of tracer galaxies.
In Figure \ref{remo}, we plot $R-I$ vs. $I-[3.6]$ colors. As noted by \citet{dolley} stars have a deficiency in $[3.6]$ $\mu$m flux compared to galaxies and can be easily separated from extended objects. 

In order to select neighbor tracer galaxies in the fields of obscured and unobscured AGN targets, we adopted the color cut $R-I<(I-[3.6])+1$.
We also plot the color of stars found in the BzK diagram (see Figure \ref{bzk}). We find a general agreement with both photometric methods.

\begin{figure}
\includegraphics[width=90mm]{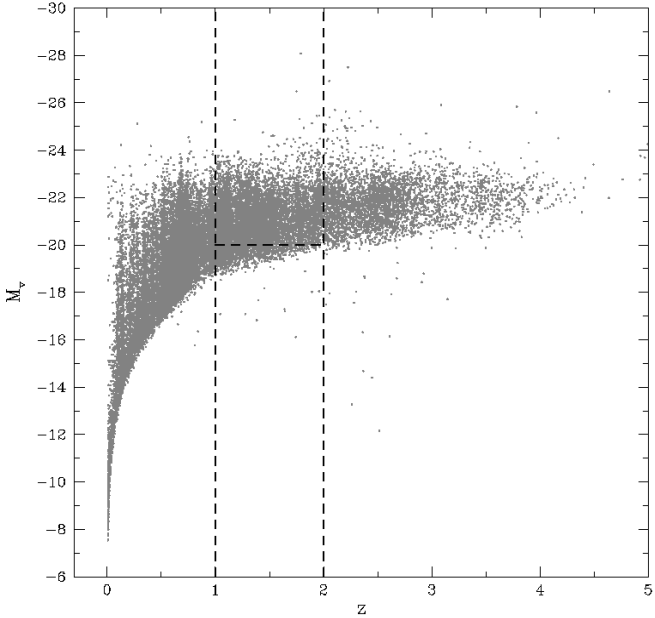}
\caption{Sample definition of our volume-limited sample in the absolute V-magnitude vs. redshift for tracer galaxies with a limiting magnitude $R_{AB}\leq$ 26.25, without stars detected using the BzK diagram. Vertical lines are the redshift cuts at z=1 and z=2 described in the text. Horizontal line corresponds to $M_v=-$20.0.}
\label{zMv}
\end{figure}

Figure \ref{zMv} shows the rest-frame $M_v$ absolute magnitude in V-band vs. redshift (spectroscopic or photometric) for our tracer galaxy sample with $R_{AB}\leq$ 26.25 and without stars detected using the BzK diagram (50124 galaxies). 
We defined a complete volume-limited sample of galaxies by selecting bounding redshifts which correspond to a specific luminosity range.
We divided the tracer galaxy sample in the redshift range of $1\leq z \leq2$ for $M_v\leq-$20.

\begin{figure}
\includegraphics[width=80mm]{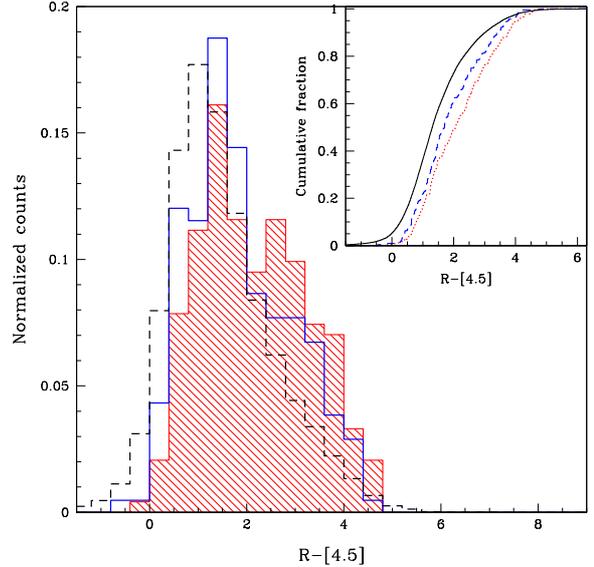}
\caption{R-[4.5] color for galaxies with $|z_{AGN}-z_{galaxy}|\leq0.2$ and $M_v\leq-20$ within 200 kpc from obscured (shade histogram) and unobscured (solid line histogram) AGNs. Dotted line histogram represent the corresponding color of the general galaxy population with  $M_v\leq-20$. The inset shows the cumulative fraction distributions.}
\label{colora}
\end{figure}

\begin{figure}
\includegraphics[width=80mm]{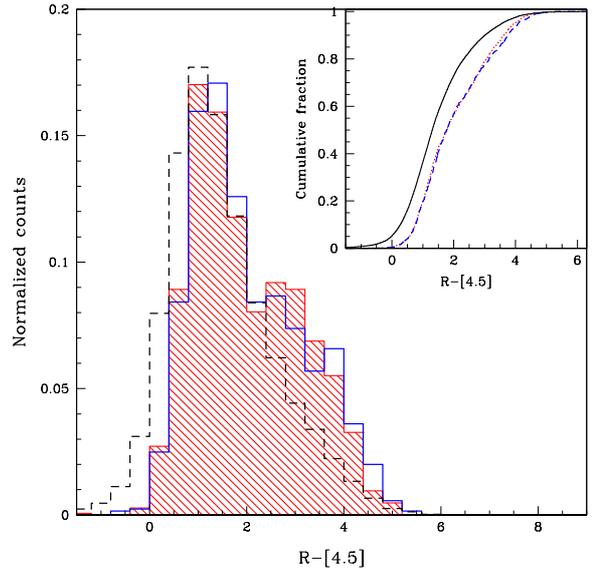}
\caption{Same as Figure \ref{colora} but for galaxies within 500 kpc. The inset shows the cumulative fraction distributions.}
\label{colorb}
\end{figure}

In Figure \ref{colora} and \ref{colorb} we present the $R-[4.5]$ color of neighbor tracer galaxies with $|z_{AGN}-z_{galaxy}|\leq0.2$ and $M_v\leq-20$ within 200 and 500 kpc, respectively from the obscured and unobscured AGN targets. The corresponding color of the general galaxy population with $M_v\leq-20  $ is shown in the figure. 
In order to make clearer the differences between the two distributions, we have included the cumulative fraction distribution for these samples. 
Small-scale environments around AGN reveals that neighbor galaxies in the field of obscured AGNs tend to have red $R-[4.5]$ colors compared with the galaxy environment of unobscured AGNs, whereas the color of tracer galaxies at larger scales ($r_p<500$ kpc) show no difference between the sample of obscured and unobscured AGN targets. 

We have also performed tests using other redshift limits, changing the redshift difference cut to $|z_{AGN}-z_{galaxy}|\leq0.1$ does not influence the trend, although the number of tracer galaxies is reduced.

We applied a Kolmogorov-Smirnov (KS) test on the color distribution of tracer galaxies in the field of obscured and unobscured AGN sample.
A KS test comparing the color distribution of galaxies within $r_p<200$ kpc from the AGN samples shows that the two distributions are different, with only a 4.3\% chance to be equal (see Figure \ref{colora}). On the other hand, the KS test applied to the sample of tracer galaxies within $r_p<500$ kpc, returned a probability of 41\%, indicating that there is no significant difference between the two distributions. This result implies that at larger scales the color of tracer galaxies around the selecting sample of AGNs are similar.

\section{Obscured and unobscured AGN environments}

\begin{figure}
\includegraphics[width=80mm]{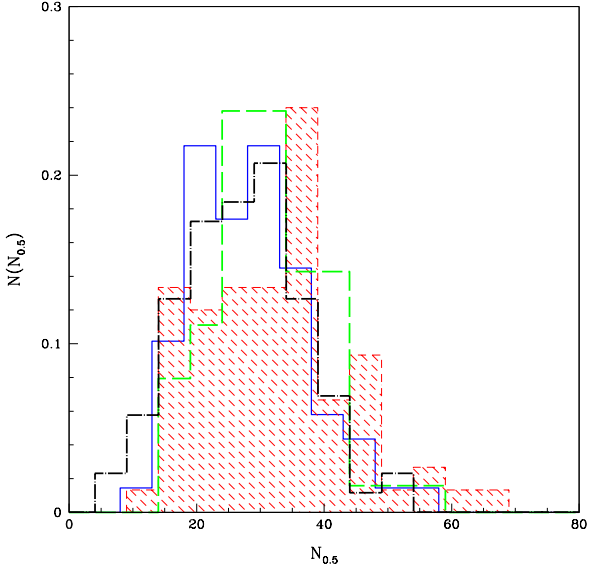}
\caption{Projected density of galaxies within 0.5 Mpc and with $|z_{AGN}-z_{galaxy}|\leq0.2$ from the obscured and unobscured AGN (shaded and solid line histograms, respectively). The dot-long dash line histogram shows the galaxy counts obtained in a control galaxy field. Long dash line histogram correspond to the distribution of galaxies in the field of Lyman-break galaxies with spectral type obtained from a star-burst galaxy. We introduce an artificial shift in the dot-long dash histogram for clarity.}
\label{n05}
\end{figure}

The study of the environments of different types of AGNs can put constraints on standard unified models.
In Figure \ref{n05} we plot the projected galaxy counts in a circular region of $0.5$ Mpc ($N_{0.5}$) with $|z_{AGN}-z_{galaxy}|\leq0.2$ around each sample of obscured and unobscured AGN. We exclude from this analysis those AGNs located in the edges of the image or in the vecinity of bright saturate stars\footnote{The sample of fits images can be found in http://www.astro.yale.edu/MUSYC/ecdfs/optical/}.
We also plot the corresponding $N_{0.5}$ values obtained for a control sample composed by random selected galaxies with redshifts and absolute V-band distributions similar to those found in AGN targets. 
The figure shows the $N_{0.5}$ distribution of galaxies using the same redshift and magnitude cuts, as explained before, in the field of a sample of Lyman-break galaxies selected by the Galaxy Evolution Explorer (GALEX) at UV wavelengths \citep{chen}. We chose a subsample of Lyman-break galaxies with spectral features of a burst galaxy and a similar redshift distribution ($1\la z\la1.4$) of obscured and unobscured AGNs. 
As seen in the figure, unobscured AGNs and Lyman-break galaxies have a similar $N_{0.5}$ distribution values than those found in control samples, while obscured AGNs are found in a dense or moderately dense environments. 

\begin{figure}
\includegraphics[width=80mm]{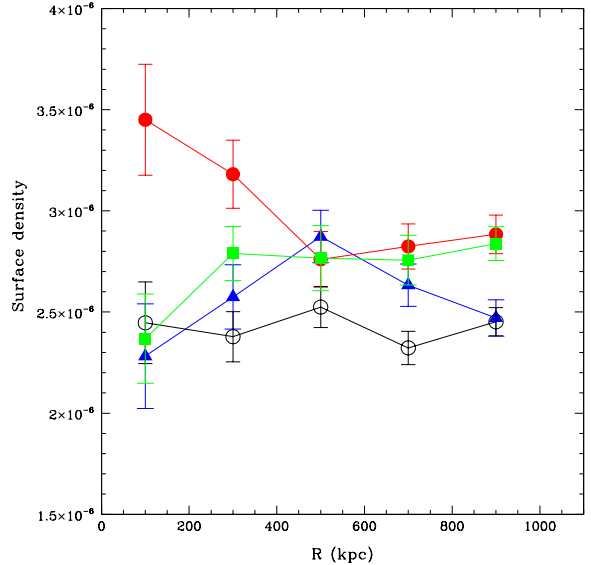}
\caption{Projected radial density of galaxies with $|z_{AGN}-z_{galaxy}|\leq0.2$  around the selected obscured and unobscured AGN (filled circles and triangles). Filled squares correspond to the density of galaxies in the field of Lyman-break galaxies. Open dots represent the corresponding values obtained in a control galaxy sample. The error bars represent the standard deviation within each data bin, estimated using Poissonian errors. }
\label{pro2}
\end{figure}

Following the results obtained in Figures \ref{colora} and \ref{colorb}, we have studied the radial density distribution of red ($R-[4.5]>2$) neighbor tracer galaxies in the field of the selected sample of AGNs.
In Figure \ref{pro2} we show the projected radial density of tracer galaxies with $|z_{AGN}-z_{galaxy}|\leq0.2$  around the obscured and unobscured AGNs with $r_p<1$ Mpc. Error bars were estimated using Poissonian errors. We excluded those AGNs located in the edges of the image or in the vecinity of bright saturate stars.
We also plot the corresponding radial profile of the control galaxy sample (open dots). 
Filled squares correspond to the radial density of galaxies in the field of Lyman-break galaxies with spectral signatures of a star-burst galaxy.
We find that red galaxies ($R-[4.5]>2$) in the field of obscured AGNs have a similar radial distribution than those of Lyman-break galaxies, while unobscured AGNs and the control sample have a similar projected radial galaxy density.

We also consider the possible existence of a selection effect of the different AGN samples since color and magnitude cuts could introduce a bias for obscured AGNs residing in more massive host galaxies compared to unobscured AGNs. However, previous works have shown that stellar masses and/or bolometric luminosities do not depend on the AGN type. For instance, \citet{hickox11} calculated the bolometric luminosity for 
samples of obscured and unobscured AGNs finding their distribution  similar. This result strongly suggest that both AGN samples have similar stellar mass.  Consistent results were given by \citet{bongio} who explored host galaxy properties of a large sample of obscured and unobscured AGNs selected according to their X-ray luminosities, spectra and multi-wavelength SEDs. For this sample, these authors also find that the stellar mass distribution of obscured and unobscured AGN is nearly independent of the AGN type.
Our results along with color distribution of neighbor tracer galaxies in the fields of AGNs, indicate that obscured AGNs are located in denser environment of evolved red galaxies compared to the environment of unobscured AGNs.

\section{Conclusions}

In this paper we analyze the properties of obscured and unobscured AGNs identified on the basis of their colors in the MIR, using the color-color criterion of \citet{stern05} and a simple criterion based on the observed optical to MIR color cut ($R -[4.5]$).
We find that the distribution of $R -[4.5]$ color is bimodal for a sample of AGNs with 1 $\leq$ z $\leq$ 2 and rest-frame absolute magnitudes $M_v\leq -$21. We obtained a similar color cut in the optical to MIR color to separate both AGN samples as proposed by \citet{hickox07}.

In agreement with other studies, we find that obscured AGNs are intrinsically optically faint in the $R$ band, suggesting that luminous IR-selected AGNs have a significant dust extinction.
The derived {\tt CLASS\_STAR} parameters suggest that the majority of obscured AGN sample have an extended emission, while unobscured AGNs have a mixture of extended and point-like emission, suggesting extended emission from the host galaxies, instead of point-like emission dominated by the nucleus.

We investigate the emission and properties of AGNs in the X-ray wavelengths. We find that X-ray emitting AGNs show a wide range of X-ray luminosity values, similar to those found in normal galaxies to those found in Seyfert-like galaxies.
We study the color distribution of galaxies in the fields of both AGN samples. We find that at small scales ($r_p<$ 200 kpc), tracer galaxies surrounding the sample of obscured AGNs tend to have redder optical to MIR colors ($R-[4.5]$) in comparison to the observed color distribution found in the unobscured AGN sample and in the control galaxy sample. Based on a KS test we find that the two color distributions are different with a $\sim$95\% significance.

Using the projected galaxy counts in a region of $0.5$ Mpc, and the projected radial density of galaxies, we find that obscured AGN are located in a denser local galaxy environment compared to the unobscured AGN sample.
Our results are in agreement with previous studies at high redshift \citep{hickox11,donoso,dipompeo}.
Similar results were also obtained from a sample of low redshift type 1 (unobscured) and type 2 (obscured) AGNs taken from the SDSS survey \citep{jiang}.
These authors found that on large scales ($\sim$ 1 Mpc), both types of AGNs have similar clustering properties, but at scales smaller than 100 kpc, type 2 AGNs have significant more neighbor galaxies than type 1.

The observed differences in the environment of distinct AGN types may also suggest an evolutionary effect, where the obscuring material is found in large galactic-scale structures, possibly due to galaxy interactions. These results do not agree with the prediction in which obscuration is purely a simple orientation effect. 

\begin{acknowledgements}
We are grateful to the anonymous referee for his/her careful reading
of the manuscript and a number of comments, which improved
the the quality of this manuscript.
This work was partially supported by the Consejo Nacional de Investigaciones Cient\'{\i}ficas y T\'ecnicas (CONICET) and the Secretar\'ia de Ciencia y Tecnolog\'ia de la Universidad de C\'ordoba (SeCyT). CGB acknowledges the assistance of Virginia Grasso. CGB and MST aknowledge support from Fondo para la Investigaci\'on Cient\'ifica y Tecnol\'ogica (FONCYT) grant PICT 2012-2377.
\end{acknowledgements}


\end{document}